\documentclass{svproc}
\newcommand{\orcidID}[1]{\unskip\textsuperscript{[#1]}}

\usepackage{url}

\usepackage{cite}
\usepackage{amsmath,amssymb,amsfonts}
\usepackage{graphicx}
\usepackage{subcaption}
\usepackage{booktabs}
\usepackage{multirow}
\usepackage{array}
\usepackage{xcolor}
\usepackage{subcaption}
\usepackage{float}
\usepackage{url}
\usepackage{makecell}

\newcommand{\setR}{\mathbb{R}}

\newcommand{\graphA}{\mathcal{A}}
\newcommand{\graphD}{\mathcal{D}}
\newcommand{\graphI}{\mathcal{I}}

\begin{document}
\mainmatter              % start of a contribution
\title{Sensor placement for real-life water distribution networks, ensuring observability under uncertainties}
\titlerunning{Sensor placement for WDNs}  % abbreviated title (for running head)
\author{Jarne J.H. van Gemert\orcidID{009-0002-8973-2533}\inst{1} \and Valentina Breschi \orcidID{0000-0002-1533-7349}\inst{1}
Claudia Quintiliani\inst{2} \and Doekle R. Yntema \orcidID{0000-0002-4432-6577}\inst{3} \and Karel J. Keesman \orcidID{0000-0003-4505-8663}\inst{3},\inst{4} \and Mircea Lazar \orcidID{0000-0003-4338-9004}\inst{1}}
\authorrunning{Jarne J.H. van Gemert, et al.} % abbreviated author list (for running head)
\tocauthor{}
\institute{Control Systems Group, Dept. of Electrical Engineering, Eindhoven University of Technology, 5600MB Eindhoven, The Netherlands\\
\email{e-mail: J.J.H.v.Gemert@tue.nl}
\and
Brabant-Water, Magistratenlaan 200, 5223 MA 's-Hertogenbosch, the Netherlands
\and
Wetsus, Centre of Excellence for Sustainable Water Technology, 8900MA Leeuwarden, The Netherlands
\and
Mathematical and Statistical Methods - Biometris, Wageningen University, 6708PB Wageningen, the Netherlands}

\maketitle              % typeset the title of the contribution

\begin{abstract}
Leakage detection in water distribution networks (WDNs) is crucial for limiting water losses. However, effective detection strongly depends on sensor availability and placement. Most existing sensor placement methods rely on extensive simulations or precise knowledge of hydraulic parameters, which are often uncertain in practice. In this paper, we propose using a scalable sensor placement framework based on structural observability that depends only on the network topology and the underlying pressure-flow interconnection structure. The approach guarantees structural observability independently of hydraulic parameters, and is therefore robust to parametric uncertainties, assuming that the network topology is correctly specified. The method is computationally efficient and its performance is demonstrated on several real-life WDNs provided by Brabant-Water. While the framework ensures full network observability, the required number of sensors can be high in large and cyclic networks, which may limit applicability in real-life and motivates future work on relaxed observability guarantees for sensor placement.
\end{abstract}

\section{Introduction}
Leakage and burst detection in water distribution networks (WDNs) is crucial for minimizing water losses, especially in the context of increasing water scarcity and climate change. Even in countries with highly developed WDN infrastructure, such as the Netherlands, approximately $5\%$ of drinking water is lost annually, while losses in other European countries can exceed $20\%$ \cite{waternewseurope2023}. 
An important factor in reducing these losses is the timely detection and localization of leakages.
To support this, numerous algorithms have been developed that aim to identify and locate leaks more quickly and accurately. These methods aim to identify abnormal network behaviour and to determine the most likely leak location. However, these methods typically rely heavily on sensor data, detailed hydraulic models, or a combination of both, where the accuracy of the hydraulic models themselves often also depends on the availability and quality of sensor data. As a result, the availability and quality of sensor data play a key role in leakage detection and localization. In practice, however, sensor installation and maintenance are costly, which limits the number of sensors that can be deployed. Therefore, the effectiveness of leakage detection depends not only on the detection method itself, but also on how sensors are placed to achieve optimal and reliable monitoring performance with minimal resources \cite{annaswamy2024control}.
\subsection{Related work}
Most traditional sensor placement methods for WDNs rely on simulation-based analyses using tools such as EPANET \cite{rossman2000epanet}, where multiple leakage scenarios are used to evaluate sensor configurations \cite{wan2022literature,santos2022pressure,ROMEROBEN202254}. Although these approaches are very intuitive, as they are based on physically interpretable quantities such as pressure, flow and topology, they typically rely on exact network information such as pipe roughness and demand patterns, which may be uncertain in practice. Additionally, their performance is only assessed for the finite set of leakage scenarios explored in simulation. 
To work towards a more structured approach, control-theoretic sensor placement methods based on observability theory have been explored \cite{geelen2021optimal}, \cite{bopardikar2021randomized}. Guaranteeing observability ensures that the system’s internal states can be reconstructed from sensor measurements within a finite time. In WDNs, this means that pressures and flows can be estimated from strategically placed sensors, allowing indirect monitoring of unmeasured parts of the network. However, many of these methods again require precise knowledge of network information and historical data. Although recent work has addressed some of these limitations by accounting for a predefined set of parametric uncertainties \cite{taha2021revisiting,zhang2024functional,mousavi2020Structural}, their dependence on specific model formulations, predefined sensor sets, or historical data, can still limit their applicability in real-life WDNs.

Given these challenges, the goal of this paper is to demonstrate that structural observability-based sensor placement can guarantee observability of WDNs independently of exact numerical parameters, if the network topology and pressure-flow interconnection structure are correctly specified. 
Additionally, we show that the proposed method remains computationally effective and applicable when applied to large-scale, real-life networks.

\section{Methodology}
In this section, we first introduce the WDN setting and its dynamic model. Building on this, we derive a structural model that depends only on the WDN topology, which is then exploited to apply the observability-guaranteeing sensor placement method as proposed in \cite{van2025exploiting}, \cite{van2025scalable}. Finally, we show the robustness of this method with respect to modelling differences.

We consider a WDN consisting of $n$ hydraulic nodes (junctions, tanks, and reservoirs) and $m$ pipes, which are represented by nodes and directed edges, respectively, see Fig. \ref{fig:1}. The system states are defined as x=
$\begin{bmatrix}
q^\top&
h^\top
\end{bmatrix}^\top\in\setR^{m+n}$,
where $q(t)\in\setR^m$ denotes the pipe flows and $h(t)\in\setR^n$ denotes the piezometric heads at hydraulic nodes, affinely related to pressure. It is important to note that $m$ is not necessarily equal to $n$, for example, four junctions can be connected by as few as three or up to six pipes. In this paper, we utilize the linearized WDN model  introduced in \cite{zeng2022elastic}, given by $\dot x(t) =Ax(t)$ with
\begin{equation}\label{eq:1}
      \dot x(t)=\begin{bmatrix}\dot{q}(t)\\\dot{h}(t)\end{bmatrix}, \text{and } A= \begin{bmatrix}\!
         -L^{-1}R\text{ diag}\begin{Bmatrix}2|{q_a}|\end{Bmatrix}\!\!&\!\! L^{-1}A_{\text{inc}}^\top\!\\
         \!F^{-1}A_{\text{inc}}\!\!&\!\!-F^{-1}D\text{diag}\begin{Bmatrix}\!\frac{1}{2\sqrt{{h_a}}}\!\!\end{Bmatrix}
    \!\end{bmatrix}.
\end{equation}
The system matrix $A \in \setR^{(m+n)\times(m+n)}$ is constructed from the incidence matrix $A_{\mathrm{inc}} \in \setR^{n\times m}$ which describes the WDN topology. The diagonal matrices $L \in \setR^{m\times m}$ and $R \in \setR^{m\times m}$ contain the hydraulic inductances and resistances of the pipes, respectively, while the diagonal matrices $F \in \setR^{n\times n}$ and $D \in \setR^{n\times n}$ represent nodal capacitance and pressure-dependent demands or discharges, respectively. Although the physical parameters in $L$, $R$, $F$, and $D$ may be uncertain, the interconnection between the states is fully determined by the network topology ($A_{\mathrm{inc}}$) and therefore remains fixed under parametric uncertainty.

\section{Structural model and data-interconnection graph}
\begin{figure*}[t]
    \centering
    \begin{subfigure}[t]{0.23\textwidth}
        \centering
        \includegraphics[width=\linewidth]{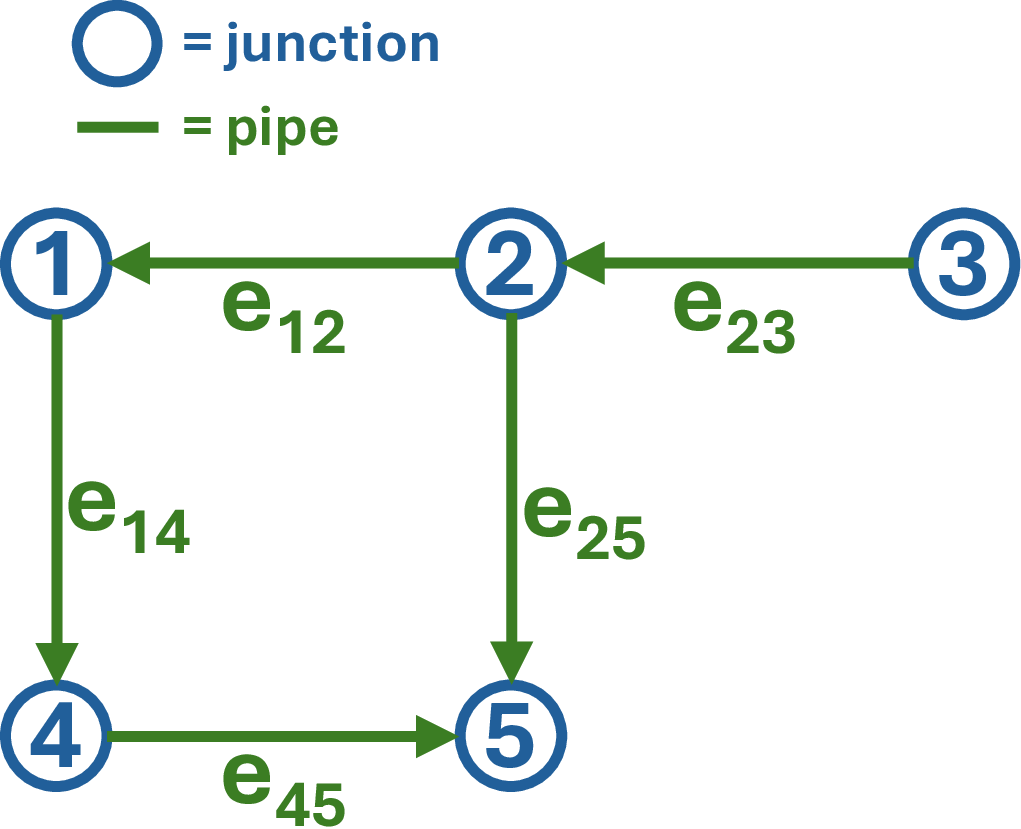}
        \caption{Topology of a WDN.}
        \label{fig:1}
    \end{subfigure}
    \hfill
    \begin{subfigure}[t]{0.23\textwidth}
        \centering
        \includegraphics[width=\linewidth]{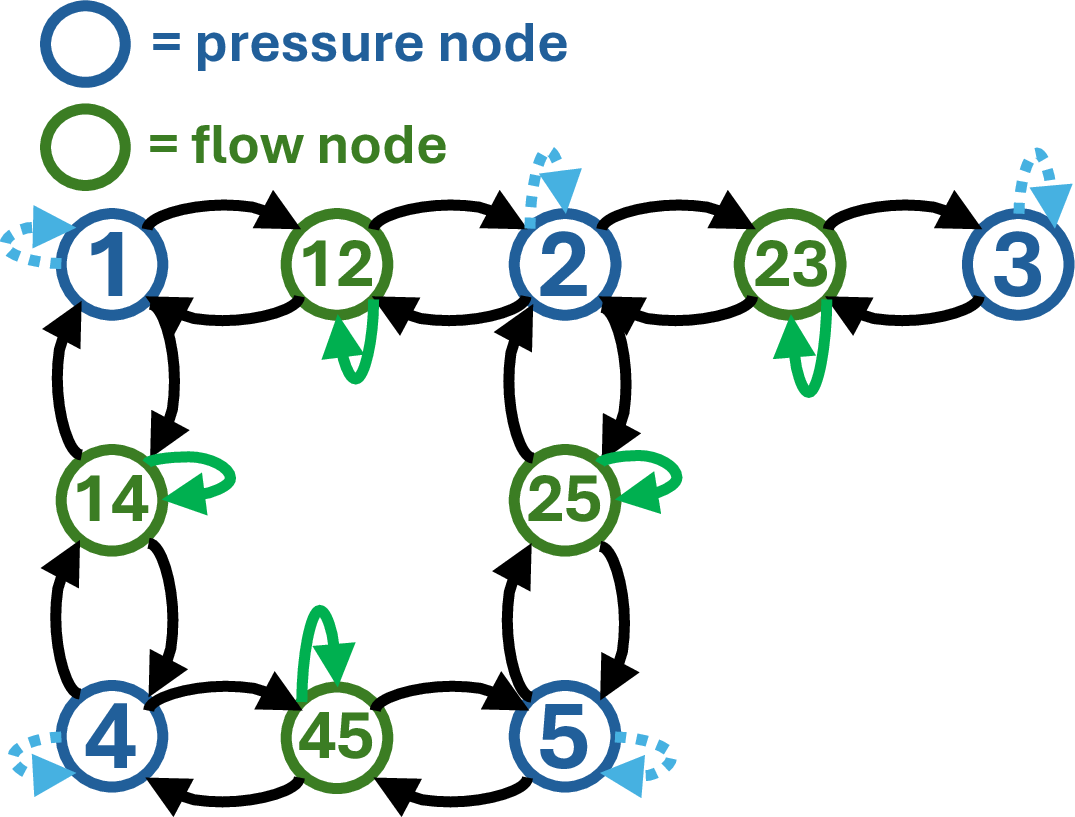}
        \caption{Data-graph based on Fig. \ref{fig:1}.}
        \label{fig:2}
    \end{subfigure}
    \hfill
    \begin{subfigure}[t]{0.23\textwidth}
        \centering
        \includegraphics[width=\linewidth]{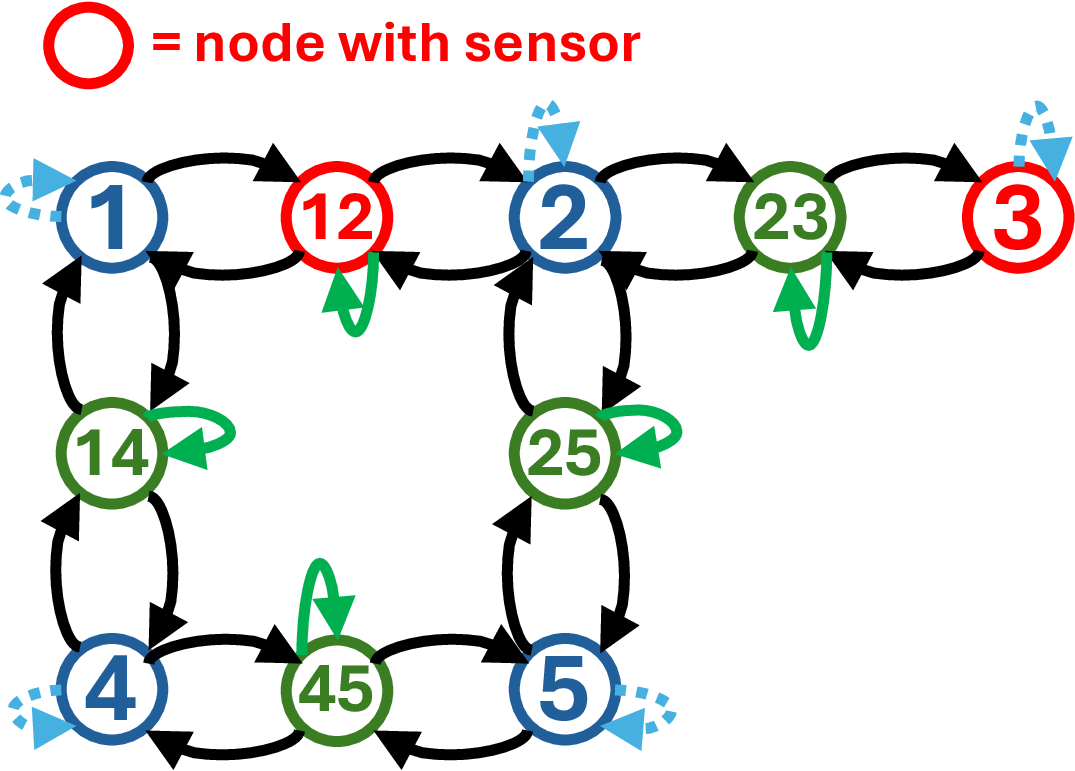}
        \caption{Colorable data-graph.}
        \label{fig:3}
    \end{subfigure}
    \hfill
    \begin{subfigure}[t]{0.23\textwidth}
        \centering
        \includegraphics[width=\linewidth]{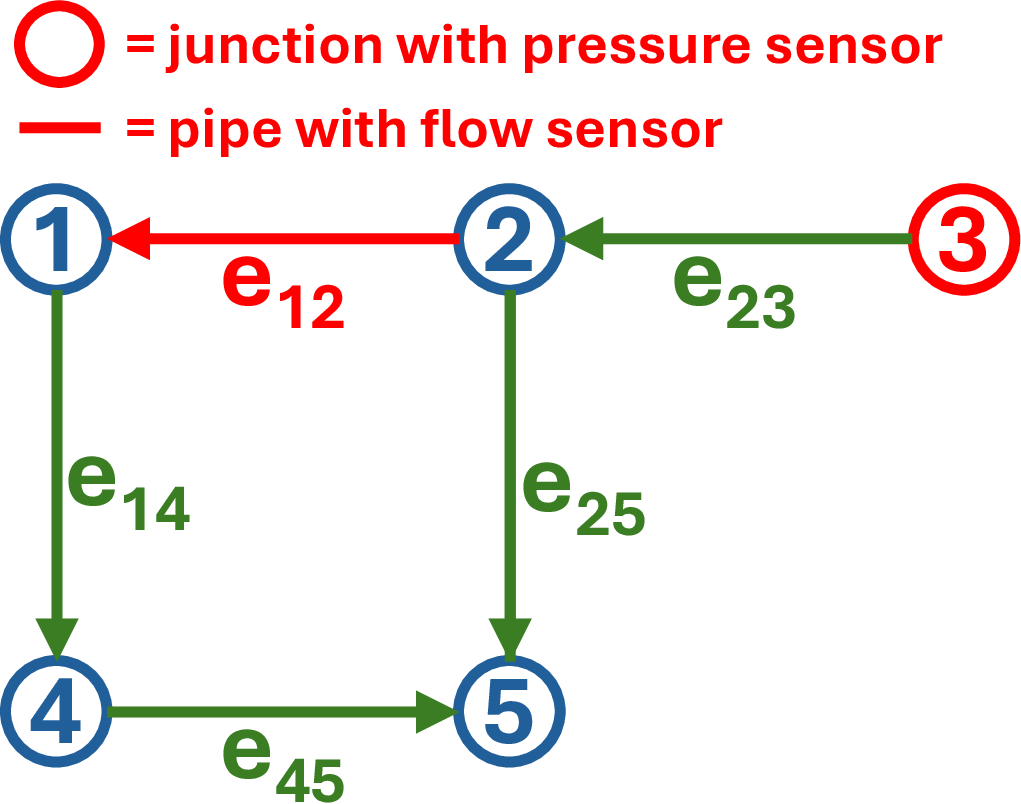}
        \caption{Sensor configuration in WDN.}
        \label{fig:4}
    \end{subfigure}
    \vspace{-0.2cm}
    \caption{Illustration of the WDN topology, its data-interconnection graph, a colorable data-interconnection-graph, and the corresponding sensor configuration.}
    \vspace{-0.5cm}
    \label{fig:overview}
\end{figure*}

Exploiting this fixed interconnection structure, we can apply structural observability to analyse and guarantee observability. By relying on structural observability rather than exact parameter values, observability is ensured for all feasible parameter realizations and for a broad class of WDN models sharing the same topological structure. To apply this method, we derive a structural representation of the linearized WDN model. This representation captures only the data-interconnection structure among states, i.e., which piezometric heads and flows can directly influence one another. This information is stored in a structural system matrix $\graphA_{\mathrm{WDN}}$, whose entries indicate the presence or absence of couplings between states: $\graphA_{\mathrm{WDN}}(i,j)=0$ if state $x_j$ never affects $x_i$, $\graphA_{\mathrm{WDN}}(i,j)=\ast$ if state $x_j$ affects $x_i$ through a nonzero coupling, and $\graphA_{\mathrm{WDN}}(i,j)=?$ if state $x_j$ might affect $x_i$, depending on operating conditions.
Based on the WDN model in \eqref{eq:1}, the structural matrix is given by
\begin{equation}
\begin{aligned}
A_{\mathrm{WDN}}&=
\begin{bmatrix}
\graphI & \graphA_{\mathrm{inc}}^\top \\
\graphA_{\mathrm{inc}} & \graphD
\end{bmatrix}, \;\text{where}\\
\graphI=\mathrm{diag}\{\ast\},\;\;\graphA_{\mathrm{inc}}(i,j)&=
\begin{cases}
0 & \text{if } \graphA_{\mathrm{inc}}(i,j)=0,\\
\ast & \text{otherwise,}
\end{cases}\;\;\graphD=\mathrm{diag}\{?\}
\label{eq:2}   
\end{aligned}
\end{equation}
Each block in $\graphA_{\mathrm{WDN}}$ represents certain physical aspects. The upper-left block $\graphI$ represents the internal pipe dynamics affecting the flow, such as resistance, which are always present in WDNs. The lower-right block $\graphD$ represents local dynamics affecting the piezometric heads, such as valve positions, which may depend on operating conditions and are therefore structurally uncertain. The off-diagonal blocks $\graphA_{\mathrm{inc}}$ describe the bidirectional coupling between flows and piezometric heads: pressure differences across connected junctions induce flows in pipes, while flows influence piezometric heads through mass balance and storage effects, which is fully determined by the network topology. For the topology in Fig. \ref{fig:1}, the incidence matrix and its corresponding structural incidence matrix are given by
\begin{equation}
A_{\mathrm{inc}}=
\begin{bmatrix}
-1 & 0 & 1 & 0 & 0\\
1 & -1 & 0 & 1 & 0\\
0 & 1 & 0 & 0 & 0\\
0 & 0 & -1 & 0 & 1\\
0 & 0 & 0 & -1 & -1
\end{bmatrix},
\quad
\graphA_{\mathrm{inc}}=
\begin{bmatrix}
\ast & 0 & \ast & 0 & 0\\
\ast & \ast & 0 & \ast & 0\\
0 & \ast & 0 & 0 & 0\\
0 & 0 & \ast & 0 & \ast\\
0 & 0 & 0 & \ast & \ast
\end{bmatrix},
\label{eq:4}
\end{equation}
with the state vector ordered as
$x(t)=
[
q_{12}(t)\, 
q_{23}(t) \,
q_{14}(t) \,
q_{25}(t) \,
q_{45}(t) \,
h_1(t) \,
h_2(t) \,$ $
h_3(t) \,
h_4(t) \,
h_5(t)
]^\top$.
Using this representation, we can obtain a data-interconnection graph by interpreting $\graphA_{\mathrm{WDN}}^\top$ as an adjacency matrix, where each state variable corresponds to a node, an $\ast$ entry results in a directed solid edge, an $?$ entry results in a directed dashed edge, and a $0$ entry indicates the absence of an edge, see Figs. \ref{fig:1} and \ref{fig:2} for an example using $A_{\mathrm{inc}}$ in \eqref{eq:4}. This graph provides a representation of how information can propagate through the WDN.

\section{Observability via graph colorability}
\begin{figure}[t]
    \centering
    \includegraphics[width=\linewidth]{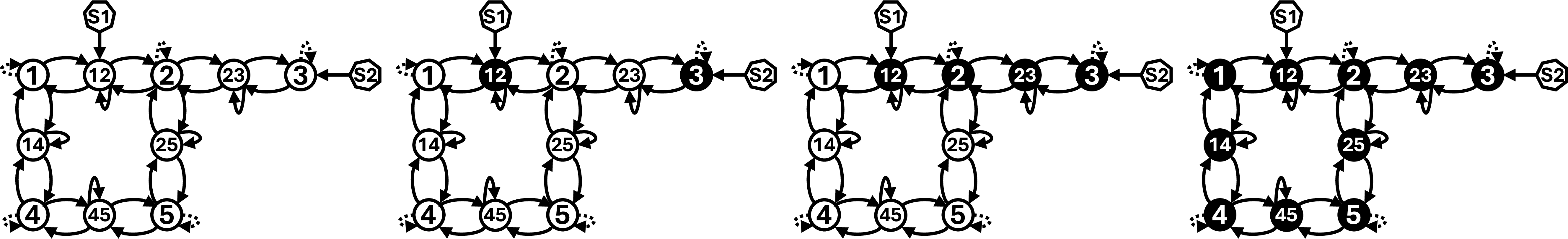}
    \caption{Example of the coloring rule for the data-interconnection graph displayed in Fig. ~\ref{fig:3}.}
    \vspace*{-0.4cm}
    \label{fig:5}
\end{figure}

To assess whether a given sensor configuration renders the WDN model structurally observable, observability is evaluated on the corresponding data-interconnection graph. Starting from the graph in Fig. \ref{fig:2}, sensor nodes are added to represent a specific sensor placement, as illustrated in Fig. \ref{fig:3}. Structural observability is then verified using a graph-based colorability rule \cite{jia2020unifying}, defined as follows. \\
\textbf{Coloring rule:} Initially, all nodes in the graph are colored white. If a node has exactly one white out-neighbor connected through a solid edge, i.e., it has exactly one outgoing solid edge to a white neighbor, while dashed edges and already colored neighbors are ignored, then this neighbor is colored. This procedure is applied iteratively, see Fig. \ref{fig:5} for an example.
Using this coloring process, the graph is said to be colorable if all state nodes, excluding the sensor nodes, are colored. 
In addition to the original data-interconnection graph, a second graph must be considered to guarantee structural observability. This graph is obtained by modifying the self-loops such that, for all non-sensor nodes missing self-loops are replaced by a solid self-loop, while existing self-loops are treated as uncertain. This results in the modified structural matrix $\bar\graphA_{\mathrm{WDN}}$ with $\bar\graphA_{\mathrm{WDN}}(i,j)=\graphA_{\mathrm{WDN}}(i,j)$ for all $i\neq j$, and
\begin{equation}\label{eq:6}
    \bar\graphA_{\mathrm{WDN}}(i,i):=\begin{cases}
\ast & \text{if } A_{\mathrm{WDN}}(i,i)=0,\\
? & \text{otherwise.}
\end{cases}
\end{equation}
If both graphs are colorable, the corresponding sensor configuration guarantees structural observability of the WDN model \cite{jia2020unifying}. However, since the data-interconnection graph of the WDN model is bidirectional, this second graph is colorable whenever the original graph is colorable \cite{van2025scalable}.
Using the colorability condition, we apply a scalable sensor placement algorithm developed in \cite[Algorithms 1-3]{van2025scalable}, which selects sensor locations such that the data-interconnection graph becomes fully colorable. Figures \ref{fig:3} and \ref{fig:4} illustrate how the resulting sensor placement translates back to the original WDN topology. 

\section{Robustness of sensor placement for modelling errors}
As shown in the previous subsections, the proposed sensor placement framework does not rely on a specific hydraulic model, but on the structural coupling between pressures and flows in the WDN. In particular, the data-interconnection graph depends only on whether state variables are coupled, e.g., the pressure-flow coupling is structurally captured by $\graphA_{\mathrm{inc}}$, and not on the exact functional form or numerical values of the underlying equations. Consequently, different WDN models that share the same diffusive pressure-flow coupling, namely,
$h_i(t)-h_j(t)=R_{ij}\left|q_{ij}(t)\right|^{\alpha-1}q_{ij}(t)$, lead to identical data-interconnection graphs, even if $R_{ij}$ and $\alpha$ differ in numerical value. I.e., as long as the pressure-flow coupling does not vanish, the corresponding entries in $\graphA_{\mathrm{WDN}}$ remain unchanged. Local effects such as pipe resistance, pressure losses, valves, and pumps appear as diagonal terms in the system matrix and correspond to self-loops in the data-interconnection graph. These self-loops do not introduce additional interconnections between states, and as shown in \cite{van2025scalable}, do not affect the graph colorability conditions for observability. As a result, variations in local modelling assumptions or operating conditions do not affect the observability guarantees.
Thus, the proposed framework applies to a broad class of WDN models with diffusive pressure--flow coupling. The resulting sensor placement is robust to variations in modelling choices and parameter values, provided that the network topology, and therefore the coupling structure between pressures and flows, is preserved.

\section{Results}
In this section, we evaluate the applicability and scalability of the proposed structural observability-based sensor placement method on large-scale, real-life WDNs provided by Brabant Water. For comparison, we include several EPANET benchmarks to put the results in perspective with respect to network size and topological properties. All experiments were performed on a Lenovo ThinkPad with an Intel Core i7-165H vPro, 64GB RAM, running MATLAB R2023a.
The method is applied to three large-scale Brabant Water networks, with the results summarized in Table \ref{tab:1}, including the network size (equal to the number of states), the number of sensors, the resulting sensor coverage (number of sensors compared to network size), and the split between flow and pressure sensors. A key observation is that the computation time scales approximately affinely with network size. This confirms that the proposed graph-based approach is suitable for real-life WDNs with tens of thousands of states. In contrast, the number of required sensors and the sensor coverage do not depend on the network size.
\begin{table}[t]
\centering
\scriptsize
\renewcommand{\arraystretch}{0.85}
\caption{Results sensor placement on real-life WDNs.}
\label{tab:1}
\begin{tabular}{lcccccc}
\hline
Network & Nodes & Sensors & \makecell{Sensor\\coverage} & \makecell{Flow\\sensors} & \makecell{Pressure\\sensors} & \makecell{Computation\\time [s]} \\
\hline
Net1 & 4095 & 226 & 5,52\% & 38 & 188 & 0,15 \\
Net2 & 59032 & 3988 & 6,76\% & 861 & 3127 & 45,92 \\
Net3 & 132893 & 9380 & 7,06\% & 2120 & 7260 & 132,46 \\
\hline
\end{tabular}
\vspace{-0.4cm}
\end{table}
To further investigate this behaviour,we compare with several EPANET benchmark networks shown in Table \ref{tab:2}. These results show that sensor coverage is not determined by network size, but rather by topological properties, most notably by cycles. Networks with a higher cycle coverage generally require more sensors to resolve flow directions within cycles, which is critical for determining information propagation and, consequently, the ability to reconstruct unmeasured states. This behaviour is consistent across both the Brabant Water networks and the EPANET benchmarks. Therefore smaller networks may require a similar sensor coverage compared to larger networks.
\begin{table}[t]
\centering
\scriptsize
\renewcommand{\arraystretch}{0.7}
\caption{Results sensor placement algorithm on EPANET benchmarks.}
\label{tab:2}
\resizebox{0.8\columnwidth}{!}{%
\begin{tabular}{lcccccc}
\hline
Network & Nodes & Sensors & \makecell{Sensor\\coverage} & \makecell{Cycle\\coverage} & \makecell{Extreme node\\coverage} & \makecell{Computation\\time [s]} \\
\hline
Hanoi & 66 & 6 & 9,1\% & 4,55\% & 4,55\% & 0,029 \\
AnyTown & 71 & 24 & 33,8\% & 26,76\% & 2,81\% & 0,030 \\
Net3 & 216 & 39 & 18,06\% & 10,64\% & 7,41\% & 0,033 \\
D-Town & 866 & 131 & 15,13\% & 6,12\% & 9,01\% & 0,060 \\
L-Town & 1694 & 162 & 9,56\% & 7,32\% & 2,18\% & 0,099 \\
\hline
\end{tabular}%
\vspace*{-0.5cm}
}
\end{table}
\section{Discussion}
In this paper, we proposed a scalable sensor placement framework for water distribution networks (WDNs) based on structural observability, relying only on the network topology. By exploiting a graph-based formulation, the method guarantees observability under modelling and parametric uncertainties and remains computationally feasible for large-scale, real-life networks.

Despite these theoretical guarantees, the practical applicability of the proposed method may be limited in real-life WDNs. In large networks, the resulting number of required sensors can be substantially high, and installing hundreds or even thousands of sensors is often not feasible for water utilities. These installations are constrained not only by financial considerations, but also by physical accessibility of installation locations, operational constraints, and maintenance requirements. Moreover, while structural observability guarantees that all pressures and flows can be reconstructed within finite time, such a strong requirement may not be necessary for leakage detection and localization. In practice, partial observability or weaker reconstruction guarantees may already suffice for reliable detection and localization of leakages, potentially offering a more realistic trade-off between monitoring performance and sensor deployment.

\vspace{-0.2cm}
\paragraph*{Acknowledgements:}
{\footnotesize\textit{
This work was performed in the cooperation framework of Wetsus, European Centre of Excellence for Sustainable Water Technology (\url{www.wetsus.nl}). Wetsus is co-funded by the European Union (Horizon Europe, LIFE, Interreg and EDRF), the Province of Fryslân and the Dutch Government: Ministry of Economic Affairs (TTT, SBO \& PPS-I/TKI Water Technology), Ministry of Education, Culture and Science (TTT \& SBO) and Ministry of Infrastructure and Water Management (National Growth Fund - UPPWATER). The authors like to thank the participants of the research theme “Smart Water Grids” for the fruitful discussions and their financial support.
}}

\bibliographystyle{spmpsci}
\bibliography{main}

\end{document}